\documentclass[10pt, conference, compsocconf]{IEEEtran}

\usepackage{ifpdf}

\ifCLASSINFOpdf
\usepackage[pdftex]{graphicx}
\else
\usepackage[dvips]{graphicx}
\fi
\usepackage[cmex10]{amsmath}
\newtheorem{definition}{Definition}

\usepackage{algorithm}
\usepackage[noend]{algorithmic}
\begin{document}
%
\title{Generalized Louvain method for community detection in large networks}


\author{
\IEEEauthorblockN{Pasquale De Meo\IEEEauthorrefmark{1}, 
									Emilio Ferrara\IEEEauthorrefmark{4}, 
									Giacomo Fiumara\IEEEauthorrefmark{1}, 
									Alessandro Provetti\IEEEauthorrefmark{1},\IEEEauthorrefmark{7}}
\IEEEauthorblockA{\IEEEauthorrefmark{1}Dept. of Physics, Informatics Section. \IEEEauthorrefmark{4}Dept. of Mathematics. University of Messina, Italy.\\
									\IEEEauthorrefmark{7}Oxford-Man Institute, University of Oxford, UK.\\
									\{pdemeo, eferrara, gfiumara, ale\}@unime.it}
}


%


\maketitle

\begin{abstract}
In this paper we present a novel strategy to discover the community structure of (possibly, large) networks.
This approach is based on the well-know concept of network modularity optimization. 
To do so, our algorithm exploits a novel measure of edge centrality, based on the $\kappa$-paths.
This technique allows to efficiently compute a edge ranking in large networks in near linear time.
Once the centrality ranking is calculated, the algorithm computes the pairwise proximity between nodes of the network.
Finally, it discovers the community structure adopting a strategy inspired by the well-known state-of-the-art \emph{Louvain method} (henceforth, LM), efficiently maximizing the network modularity.
The experiments we carried out show that our algorithm outperforms other techniques and slightly improves results of the original LM, providing reliable results.
Another advantage is that its adoption is naturally extended even to unweighted networks, differently with respect to the LM.
\end{abstract}


%
\IEEEpeerreviewmaketitle

\section{Introduction}
The investigation of the community structure inside networks has acquired a great relevance during the last years, in particular in the context of Social Network Analysis (SNA).
This, also because of the unpredicted success of Online Social Networks (OSNs).
In fact, social phenomena such as Facebook and Twitter amongst others, glue together millions of users under a unique network whose features are a goldmine for Social Scientists.
Several works are focused on the Social Network analysis of these OSNs; others describe the strategies of analysis themselves.

In this paper we focus on the possible strategies of community detection.
As to date, two paradigms exist to discover the community structure of a network.
The former is based on the analysis of the global features of the network, for example its topology. 
These approaches are characterized by high computational complexity and high quality results.
The latter paradigm relies on exploiting local information, for example those acquirable by nodes and their neighborhoods. 
The computational cost of these techniques is lower than those exploiting global features, but the reliability decreases.

In this work, we propose a novel strategy to discover the inner community structure of a network.
The main characteristics of our approach are the followings: i) it exploits global information of the network, establishing which are the edges of the network that contribute to the creation of the community structure; ii) to do so, it adopts a novel measure of edge centrality, in order to rank all the edges of the network with respect to their proclivity to propagate information through the network itself; iii) its computational cost is low, making it feasible even for large network analysis; iv) it is able to produce reliable results, even if compared with those calculated by using more complex techniques, when this is possible; in fact, because of the computational constraints, the adoption of some existing techniques is not viable when considering large networks, and their application is only limited to small case-studies.

This paper is organized as follows: in the next Section we provide some background information about the \emph{community detection} problem.
Section \ref{sec:design-goals} introduces the main objectives of this work and describes an intuitive sketch about the novel strategy of community detection we propose.
In Section \ref{sec:kpath-centrality} the key concept of $\kappa$-path edge centrality is recalled, being it a novel and efficient strategy of ranking edges with respect to their centrality in the network.
All the pieces are glued together in Section \ref{sec:community-detection}. 
We describe our strategy to detect the community structure, inspired by the well-known state-of-the-art LM \cite{blondel2008fast}, which is computationally suitable even when large networks are analyzed.
Experiments that have been carried out are discussed in Section \ref{sec:experimentation}.
Finally, Section \ref{sec:conclusions} concludes, depicting some future directions of research.

\section{Background} \label{sec:background}
Several techniques to investigate the community structure of networks have been proposed in literature during last years. 
There exist numerous comprehensive surveys to this problem, such as \cite{porter2009communities,fortunato2010community}.

In its general formulation, the problem of finding communities in a network is intended as a data clustering problem.
In fact, it could be solved assigning each node of the network to a cluster, in a meaningful way.
Two approaches have been widely investigated, i) \emph{spectral clustering} based techniques, and, ii) \emph{network modularity optimization} strategies.
The former relies on the optimization of the process of cutting the graph representing the given network.
The latter is based on the maximization of a benefit function, called \emph{network modularity}.
We briefly recall them, separately.

The problem of minimizing the number of cuts in a given graph has been proved to be NP-hard.
To do so, different approximate techniques have been proposed.
An example is by using the spectral clustering \cite{ng2001spectral}, exploiting the eigenvectors of the Laplacian matrix of the network.
We recall that the Laplacian matrix $L$ of a given graph has components $L_{ij} = k_i\delta(i,j)-A_{ij}$, where $k_i$ is the degree of a node $i$, $\delta(i,j)$ is the Kronecker delta (that is, $\delta(i,j)= 1$ if and only if $i=j$) and $A_{ij}$ is the adjacency matrix representing the graph connections.
Another approach relies on the strategy of the ratio cut partitioning \cite{wei1989towards,hagen2002new}.
This is a function that, if minimized, allows the identification of large clusters with a minimum number of outgoing interconnections. 
The principal issue of spectral clustering based techniques is that one has to know in advance the number and the size of communities comprised in the given network.
This makes this strategy unfeasible if the purpose is to discover the unknown community structure of a network.

The strategy exploited in this paper adopts the second paradigm, the one relying on the concept of \emph{network modularity}.
It can be explained as follows: let consider a network, represented by means of a graph $G=(V,E)$, partitioned into $m$ communities; assuming $l_s$ the number of edges between nodes belonging to the $s$-th community and $d_s$ is the sum of the degrees of the nodes in the $s$-th community, the network modularity $Q$ is given by

\begin{equation}
	\label{eq:qmod}
	Q= \sum_{s = 1}^m \left[\frac{l_s}{|E|} - \left(\frac{d_s}{2|E|}\right)^2\right]
\end{equation} 

Intuitively, high values of $Q$ implies high values of $l_s$ for each discovered community; thus, detected communities are dense within their structure and weakly coupled among each other.
Equation \ref{eq:qmod} reveals a possible maximization strategy: in order to increase the value of the first term (namely, the \emph{coverage}), the highest possible number of edges should fall in each given community, whereas the minimization of the second term is obtained by dividing the network in several communities with small total degrees.

The problem of maximizing the network modularity has been proved to be NP complete \cite{brandes2007finding}.
To this purpose, several heuristic strategies to maximize the network modularity $Q$ have been proposed as to date.
Probably, the most popular one is called \emph{Girvan-Newman strategy} \cite{girvan2002community,newman2004finding}.
This approach works in two steps, i) ranking edges by using the betweenness centrality as measure of importance; ii) deleting edges in order of importance, evaluating the increase of the value of $Q$. 
In fact, it is possible to maximize the network modularity deleting edges with high value of betweenness centrality, based on the intuition that they connect nodes belonging to different communities.
The process iterates until a significant increase of $Q$ is obtained.
At each iteration, each connected component of $S$ identifies a community.
Unfortunately, the computational cost of this strategy is very high (i.e., $O(n^3)$, being $n$ the number of nodes in $S$).
This makes it unsuitable for the analysis of large networks.
The largest part of its cost is given by the calculation of the betweenness centrality, that is itself very costly (even if the most efficient algorithm \cite{brandes2001faster} is adopted).

Several variants of this strategy have been proposed during the years, such as the fast clustering algorithm provided by Clauset, Newman and Moore \cite{clauset2004finding}, that runs in $O(n \log n)$ on sparse graphs; the extremal optimization method proposed by Duch and Arenas \cite{duch2005community}, based on a fast agglomerative approach, with $O(n^2 \log n)$ time complexity; the Newman-Leicht \cite{newman2007mixture} mixture model based on statistical inferences; other maximization techniques by Newman \cite{newman2006finding} based on eigenvectors and matrices.

The state-of-the-art technique is called \emph{Louvain method} (LM) \cite{blondel2008fast}.
This strategy is based on local information and is well-suited for analyzing large weighted networks.
It is based on the two simple steps: i) each node is assigned to a community chosen in order to maximize the network modularity $Q$; the gain derived from moving a node $i$ into a community $C$ can simply be calculated as \cite{blondel2008fast} 

\begin{equation}
\tiny
	\Delta Q= \frac{\sum_{C} + k_i^C}{2m} - \left( \frac{\sum_{\hat{C}} + k_i}{2m} \right)^2 
	-\left[ \frac{\sum_{C}}{2m} - \left( \frac{\sum_{\hat{C}}}{2m} \right)^2 - \left( \frac{k_i}{2m} \right) \right]	
\label{eq:deltaq}
\end{equation}

where $\sum_{C}$ is the sum of the weights of the edges inside $C$, $\sum_{\hat{C}}$ is the sum of the weights of the edges incident to nodes in $C$, $k_i$ is the sum of the weights of the edges incident to node $i$, $k_i^C$ is the sum of the weights of the edges from $i$ to nodes in $C$, $m$ is the sum of the weights of all the edges in the network;
ii) the second step simply makes a new network consisting of nodes that are those communities previously found. 
Then the process iterates until a significant improvement of the \emph{network modularity} is obtained.

In this paper we present an efficient community detection algorithm which represents a generalization of the LM.
In fact, it can be applied even on unweighted networks and, most importantly, it exploits both global and local information.
To make this possible, our strategy computes the pairwise distance between nodes of the network.
To do so, edges are weighted by using a global feature which represents their aptitude to propagate information through the network.
The edge weighting is based on the $\kappa$-path edge centrality, a novel measure whose calculation requires a near linear computational cost \cite{cikm2011}.
Thus, the partition of the network is obtained improving the LM.
Details of our strategy are explained in the following.

\section{Design Goals} \label{sec:design-goals}
In this Section we briefly and informally discuss the ideas behind our strategy. 
First of all, we explain the principal motivations that make our approach suitable, in particular but not only, for the analysis of the community structure of Social Networks. 
To this purpose, we introduce a real-life example from which we infer some features of our approach.

Let consider a social network, in which users are connected among them by friendship relations. 
In this context, we can assume that one of the principal activities could be exchanging information.
Thus, let assume that a ``message'' (that, could be, for example, a \emph{wall post} on Facebook or a \emph{tweet} on Twitter) represents the simplest ``piece'' of information and that users of this network could exchange messages, by means of their connections. 
This means that a user could both directly send and receive information only to/from the people in her neighborhood.
In fact, this assumption will be fundamental (see further), in order to define the concepts of \emph{community} and \emph{community structure}.
Intuitively, say that a community is defined as a group of individuals in which the interconnections are denser than outside the group (in fact, this maximizes the benefit function $Q$).

The aim of our community detection algorithm is to identify the partitioning of the network in communities, such that the network modularity is optimal.
To do so, our strategy is to rank links of the network on the basis of their aptitude of favoring the diffusion of information.
In detail, the higher the ability of a node to propagate a message, the higher its centrality in the network.
This is important because, as already proved by \cite{girvan2002community,newman2004finding}, we could ensure that the higher the centrality of a edge, the higher the probability that it connects different communities.

Our algorithm adopts different optimizations in order to efficiently compute the link ranking.
Once we define an optimized strategy for ranking links, we can compute the pairwise distances between nodes and finally the partitioning of the network, according to the LM.
The evaluation of the goodness of the partitioning in communities is attained by adopting the measure of the network modularity $Q$.

In the next sections we shall discuss how our algorithm is able to incorporate these requirements.
First of all, in Section \ref{sec:kpath-centrality}, we formally provide a definition of centrality of edges in social networks based on the propagation of messages by using simple random walks of length at most $\kappa$ (called, hereafter, $\kappa$-{\em path edge centrality}).
Then, we provide a description of an efficient implementation of this algorithm, running in $O(\kappa |E|)$, where $|E|$ is the number of edges in the network.
After this, in Section \ref{sec:community-detection} we discuss the technical details of our community detection algorithm.

\section{{\large $\kappa$}-Path Edge Centrality} \label{sec:kpath-centrality}
The concept of $\kappa$-path edge centrality has been recently defined \cite{cikm2011} as follows:

\smallskip
\begin{definition} ($\kappa$-path edge centrality) For each edge $e$ of a graph $G = (V,E)$, the $\kappa$-path edge centrality $L^\kappa(e)$ of $e$ is defined as the sum, over all possible source nodes $s$, of the percentage of times that a message originated from $s$ traverses $e$, assuming that the message traversals are only along random simple paths of at most $\kappa$ edges.
\end{definition}

The \emph{$\kappa$-path edge centrality} is formalized, for an arbitrary edge \emph{e}, as follows
\begin{equation}
L^\kappa(e) = \sum_{s \in V}{\frac{\pi_s^\kappa(e)}{\pi_s^\kappa}}
\label{eq:edge-k-path}
\end{equation}

where $s$ are all the possible source nodes, $\pi_s^\kappa(e)$ is the number of $\kappa$-paths originating from $s$ and traversing the edge $e$ and $\pi_s^\kappa$ is the number of $\kappa$-paths originating from $s$.

\subsection{Fast $\kappa$-path Edge Centrality Algorithm} \label{sec:edge-k-path-centrality}
In this section we recall the functioning of the strategy adopted to efficiently compute the $\kappa$-path edge centrality.
The proposed algorithm \cite{cikm2011} is called \emph{Weighted Edge Random Walk $\kappa$-Path Centrality} (or, shortly, WERW-Kpath).
It consists of two main steps: i) node and edge weights assignment, and ii) simulation of message propagations using random simple paths of length at most $\kappa$.
In the following, the two steps are discussed separately.

\subsubsection{Step 1}{Node and edge weights assignment}

In the first stage, the algorithm assigns a weight to both nodes and edges of the graph $G = (V,E)$ representing the given network.
Node weights are exploited to choose the source nodes from which the simulation of the message propagations starts; edge weights represent initial values of centrality and they are updated during the execution of the algorithm.
At the end of the execution of $\rho$ simulations, where the optimal value $\rho = |E|-1$ has been proved in \cite{cikm2011}, edge weights are exploited for the edge ranking.

To compute node weights, we recall the notion of \emph{local effective density} $\delta(v)$ of a node $v$, as follows:

\smallskip
\begin{definition}{\emph{Local effective density}}
	Given a graph $G=(V,E)$ and a node $v$, its local effective density $\delta(v)$ is
	$\displaystyle{\delta(v) = \frac{|I(v)|+|O(v)|}{2|E|}}$
	where $I(v)$ and $O(v)$ represent, respectively, the number of ingoing and outgoing edges incident on the node $v$.
\end{definition}

This value intuitively represents how much a node contributes to the overall connectivity of the graph.
The higher $\delta(v)$, the better $v$ is connected in the graph.

As for edge weights, we recall the following definition:

\smallskip
\begin{definition}{\emph{Initial edge weight}}
	Given a graph $G=(V,E)$ and an edge $e$, its initial edge weight $\omega(e)^{0}$ is
	$\displaystyle{\omega(e)^{0}  = \frac{1}{|E|}}$
	where $|E|$ is the cardinality of $E$.
\end{definition}

Intuitively, we initially manage a ``budget'' consisting of $|E|$ points; these points are equally divided among all the nodes; the amount of points received by each edge represents its initial rank.

\subsubsection{Step 2}{Simulation of message propagations}

In the second step we simulate $\rho$ simple random walks of length at most $\kappa$ on the network.
In detail, at each iteration, WERW-Kpath (Algorithm \ref{alg:ERW-Kpath}) performs these operations:

\begin{enumerate}
	\item A node $v$ of the graph $G$ is selected with a probability  proportional to its local effective density $\delta(v)$
						
			\begin{equation}
				\displaystyle{P(v)=\frac{\delta(v)}{\phi}}
				\label{eq:ve}
			\end{equation}
						
			where $\displaystyle{\phi= \sum_{v \in V} \delta(v)}$ is a normalization factor.
	
	\item  All the edges in $G$ are marked as not traversed.
	
	\item The procedure {\em MessagePropagation} is invoked. 
\end{enumerate}

\begin{algorithm}
\small
	\caption{WERW-Kpath(Graph $G=(V,E)$, int $\kappa$)}
	\label{alg:ERW-Kpath}
	\begin{algorithmic}[1]
		\STATE Assign each node $v$: $\delta(v)=\frac{|I(v)|+|O(v)|}{|E|}$
		\STATE Assign each edge $e$: $\omega(e)=\frac{1}{|E|}$
		\STATE $\rho \leftarrow |E|-1$
		\FOR{$i=1$ to $\rho$}			
			\STATE $N \leftarrow 0$ a counter to check the length of the $\kappa$-path
			\STATE $v \leftarrow$ a node chosen according to Eq. \ref{eq:ve}
			\STATE MessagePropagation($v$, $N$, $\kappa$)
		\ENDFOR
	\end{algorithmic}
\end{algorithm}

Let us describe the procedure {\em MessagePropagation} (Algorithm \ref{alg:procedure}).
This procedure carries out a loop until {\em both} the following conditions hold true:

\begin{itemize}

	\item The length of the path currently generated is no greater than $\kappa$.
				This is managed through a length counter $N$.
	
	\item Assuming that the walk has reached the node $v_n$, there must exist at least an outgoing edge from $v_n$ which has not been already traversed.
				In detail, we attached a flag $T(e)$ to each edge $e$; $T(e) = 1$ if the edge $e$ has already been traversed, 0 otherwise.
				If we	call $O(v_n)$ the set of outgoing edges from $v_n$, it must hold that
				$\displaystyle{|O(v_n)| \neq \sum_{e \in O(v_n)}{T(e)}}$.
				
\end{itemize}

The former condition allows us to consider only paths up to length $\kappa$.
The latter condition, instead, avoids that the message get trapped into a cycle.

If the conditions above are satisfied, the {\em MessagePropagation} procedure selects an edge $e_n$ with a probability  proportional to the edge weight $\omega(e_n)$, given by

	\begin{equation}
		\displaystyle{P(e_n)=\frac{\omega(e_n)}{\gamma}}
		\label{eq:p-e-n}
	\end{equation}
					
	where $\displaystyle{\gamma= \sum_{e_n \in \hat{O}(v_n)} \omega(e_n)}$ is a normalization factor, 
	being $\hat{O}(v_n) = \{ e_n \in O(v_n) \ | \ T(e_n)=0 \}$.

Let $e_n$ be the selected edge and let $v_{n+1}$ be the node reached from $v_n$ by means of $e_n$.
The {\em MessagePropagation} procedure awards a bonus (equal to $\beta = \frac{1}{|E|}$) to $e_n$, sets $T(e_n) = 1$ and increases the counter $N$ by 1.
The message propagation activity continues from $v_{n+1}$.

At the end of all the processes of simulation of message propagation, each edge $e \in E$ is assigned a centrality value $L^\kappa(e)$ (in the interval $[\frac{1}{|E|},1]$) equal to its final weight $\omega(e)$.

The time complexity of this algorithm is $O(\kappa |E|)$. 
Our community detection strategy, described in the following, adopts this algorithm to weight edges of the network.

\begin{algorithm}
\small
	\caption{MessagePropagation(Node $v$, int $N$, int $\kappa$)}
	\label{alg:procedure}
	\begin{algorithmic}[1]
				\WHILE{$N < \kappa$ and $\left(|O(v)| \neq \sum_{e \in O(v)}{T(e)}\right)$}
				\STATE $e_{vw} \leftarrow e \in \{O(v)\ | \ T(e) = 0\}$, according to Eq. \ref{eq:p-e-n}
				\STATE $\omega(e_{vw}) \leftarrow \omega(e_{vw}) + \frac{1}{|E|}$
				\STATE $T(e_{vw}) \leftarrow  1$; $v \leftarrow  w$; $N \leftarrow N+1$
			\ENDWHILE
	\end{algorithmic}
\end{algorithm}

\section{Community Structure Detection} \label{sec:community-detection}
In the following, we present a novel algorithm to calculate the community structure of a network.
It is baptized \emph{Fast $\kappa$-path Community Detection} (or, shortly, FKCD).
The strategy relies on three steps: 
i) ranking edges by using the WERW-Kpath algorithm; 
ii) calculating the proximity (the inverse of the distance) between each pair of connected nodes;
ii) partitioning the network into communities so to optimize the network modularity \cite{girvan2002community}, according to the LM \cite{blondel2008fast}.
The algorithm is discussed as follows.

\subsection{Fast $\kappa$-path Community Detection}
First of all, our Fast $\kappa$-path Community Detection (henceforth, FKCD) needs a ranking criterion to compute the aptitude of all the edges to propagate information through the network.
To do so, FKCD invokes the WERW-Kpath algorithm, previously described.
Once all the edges have been labeled with their $\kappa$-path edge centrality, a ranking in decreasing order of centrality could be obtained.
This is not fundamental, but could be useful in some applications.
Similarly, before to proceed, a first \emph{network modularity} esteem (hereafter, $Q$) could be calculated.
This could help in order to put into evidence how $Q$ increases during next steps.
With respect to $Q$, we recall that its value ranges in the interval $[0,1]$ and, the higher $Q$, the better the community structure of the network appears evident.
The computational cost of this first step is $O(\kappa|E|)$, with $\kappa$ length of the $\kappa$-paths and $|E|$ cardinality of $E$.

The second step consists in calculating the proximity among each pair of connected nodes.
This is done by using a $L_2$ distance (i.e., the \emph{Euclidean distance}) calculated as
\begin{equation}
	r_{ij} = \sqrt{\sum_{k=1}^{n}{\frac{(L^{\kappa}(e_{ik}) - L^{\kappa}(e_{kj}))^2}{d(k)}}}
\label{eq:distance}
\end{equation}

where $L^{\kappa}(e_{ik})$ (resp., $L^{\kappa}(e_{kj})$) is the $\kappa$-path edge centrality of the edge $e_{ik}$ (resp., $e_{kj}$) and $d(k)$ is the degree of the node.
We put into evidence that, even though the $L_2$ measure would return a distance, in our case, the higher $L^{\kappa}(e_{ik})$ (resp., $L^{\kappa}(e_{kj})$), the more the nodes are \emph{near}, instead of distant.
This important aspect leads us to consider the results of Equation \ref{eq:distance} as the pairwise \emph{proximities} of nodes.
This step is theoretically computationally expensive, because it should require $O(|V|^2)$ iterations, but in practice, by adopting optimization techniques, its near linear cost is $O(\overline{d}(v)|V|)$, where $\overline{d}(v)$ is the mean degree of all the nodes of the network (and it is usually small in Social Networks).

The last step is the network partitioning. 
The main idea is inspired by the LM \cite{blondel2008fast} for detecting the community structure of weighted networks in near linear time.
The partitioning is an iterative process.
At each iteration, two simple steps occur: i) each node is assigned to a community chosen in order to maximize the network modularity $Q$; the possible increase of $Q$ derived from eventually moving a node $i$ into a community $C$ is calculated according with Equation \ref{eq:deltaq}; ii) the second step produces a \emph{meta-network} whose nodes are those communities previously found. 
The partitioning ends when no further improvements of $Q$ can be obtained.

This reflects in splitting communities connected by edges with high proximity, which is a global feature, thus maximizing the network modularity. 
Its cost is $O(\gamma|V|)$, where $|V|$ is the cardinality of $V$ and $\gamma$ is the number of iterations required by the algorithm to converge (in our experience, usually, $\gamma < 5$).
The FKCD is schematized in Algorithm \ref{alg:fkcd}.

We recall that \textsl{CalculateDistance} computes the pairwise node distance by using Equation \ref{eq:distance}, \textsl{Partition} extracts the communities according to the LM descripted above and \textsl{NetworkModularity} calculates the value of \emph{network modularity} by using Equation \ref{eq:qmod}.

The computational cost of our strategy is near linear.
In fact, $O(\kappa |E| + \overline{d}(e) |V| + \gamma |V|) = O(\Gamma |E|)$, by adopting efficient graph memorization in order to minimize the execution time for the computation of Equations \ref{eq:qmod} and \ref{eq:distance}.

\begin{algorithm}
\small
	\caption{FKCD(Graph $G=(V,E)$, int $\kappa$)}
	\label{alg:fkcd}
	\begin{algorithmic}[1]
		\STATE WERW-Kpath(G, $\kappa$)
		\STATE CalculateDistance(G)
		\WHILE{$Q$ increases at least of $\epsilon$ (arbitrarily small)}
		\STATE P = Partition(G)
		\STATE $Q \leftarrow$ NetworkModularity(P)
		\ENDWHILE
	\end{algorithmic}
\end{algorithm}

\section{Experimental Results} \label{sec:experimentation}
Our experimentation has been conducted both on synthetic and real-world online social networks, whose datasets are available online.
All the experiments have been carried out by using a standard Personal Computer equipped with a Intel i5 Processor with 4 GB of RAM.

\subsection{Synthetic Networks}

The method proposed to evaluate the quality of the community structure detected by using the FKCD exploits the technique presented by Lancichinetti et al. \cite{lancichinetti2008benchmark}.
We generated the same synthetic networks reported in \cite{lancichinetti2008benchmark}, adopting the following configuration: i) $N=1000$ nodes; ii) the four pairs of networks identified by $(\gamma,\beta) = (2,1),(2,2),(3,1),(3,2)$, where $\gamma$ represents the exponent of the \emph{power law distribution of node degrees}, $\beta$ the exponent of the \emph{power law distribution of the community sizes}; iii) for each pair of exponents, three values of average degree $\langle k \rangle = 15,20,25$; iv) for each of the combinations above, we generated six networks by varying the \emph{mixing parameter} $\mu = 0.1, 0.2, \dots, 0.6$.

\begin{figure}[!th]%
\centering
	\includegraphics[width=\columnwidth]{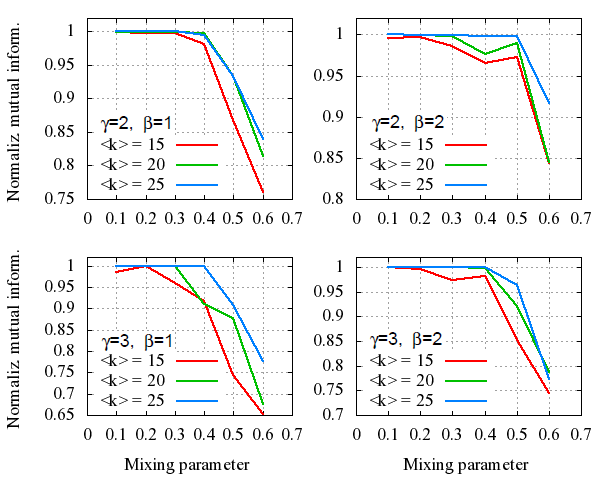}%
	\caption{Test of modularity optimization using the benchmark \cite{lancichinetti2008benchmark}, for N = 1000 nodes.
	The threshold value $\mu=0.5$ represents the border beyond which communities are no longer defined in the strong sense, i.e., each node has more neighbors in its own community than in the others \cite{radicchi2004defining}.}%
	\label{fig:1k}%
\end{figure}

Figure \ref{fig:1k} highlights the quality of the obtained results.
The measure adopted is the \emph{normalized mutual information} \cite{danon2005comparing}.
Values obtained put into evidence that our strategy performs fair good results, avoiding the well-known effect due to the resolution limit of the modularity optimization \cite{fortunato2007resolution}.
Moreover, a classification of results as in Table \ref{tab:datasets} (discussed later) is omitted because values of $Q$ obtained by using FKCD and the LM in the case of these quite small synthetic networks are very similar.

\subsection{Real-world Networks}
Results obtained by analyzing several real-world networks \cite{leskovec2006sampling,viswanath2009evolution} are summarized in Table \ref{tab:datasets}.
This experimentation has been carried out to qualitatively analyze the performance of our strategy.
Obtained results, measured by means of the \emph{network modularity} calculated by our algorithm (FKCD), are compared against those obtained by using the original LM.

Our analysis puts into evidence the following observations: i) classic not optimized algorithms (for example Girvan-Newman \cite{girvan2002community}) are unfeasible for large network analysis; ii) results obtained by using LM are slightly higher than those obtained by using FKCD; on the other hand, LM adopts local information in order to optimize the network modularity, while our strategy exploits both local and global information; this results in (possibly) more convenient identified community structures for some applications; iii) the performance of FKCD slightly increase by using longer $\kappa$-paths; iv) both the compared efficient strategies are feasible even if analyzing large networks using standard resources of calculus (i.e., a classic personal computer); this aspect is important if we consider that there exist several Social Network Analysis tools (e.g., NodeXL\footnote{http://nodexl.codeplex.com/}) that require optimized fast algorithms to compute the community structure of networks.

\begin{table}[!th]
	\centering
	\footnotesize
	\caption{Datasets adopted in our experimentation}
	\begin{tabular}{@{}c@{}c@{}c@{}c@{}c@{}c@{}c@{}}
		\hline
		Network 			&	No. nodes	\	&	\ No. edges	\	& \ No. comm. \ & 		\ Fkcd$_{\kappa=5}$ \ & \ Fkcd$_{\kappa=20}$ \ & \ Lm\\	
		\hline \hline
		CA-GrQc				&	5,242			&	28,980		& 883   &	0.734	 &	0.786  &	0.816 \\
		CA-HepTh			&	9,877			&	51,971		& 1,501 &	0.585	 &	0.648  &	0.768 \\
		CA-HepPh			&	12,008		&	237,010		&	1,243	&	0.565  &	0.598  &	0.659 \\
		CA-AstroPh		&	18,772		&	396,160		&	1,552	&	0.486  &	0.568  &	0.628 \\	
		CA-CondMat		&	23,133		&	186,932		& 2,819 & 0.546	 &	0.599  &	0.731 \\	
		Facebook			&	63,731		&	1,545,684	&	6,484	&	0.414	 & 0.444	 &	0.634	\\
		\hline
	\end{tabular}
	\label{tab:datasets}
\end{table}

\section{Conclusions} \label{sec:conclusions}
The problem of discovering the community structure in large networks has been widely investigated during last years.
Several efficient approaches based on local information have been proposed, and are feasible even when analyzing large networks because of their low computational cost.
The main drawback of the existing techniques is that they do not consider global information about the topology of the network. 
In this work we presented a novel strategy that has two advantages. 
The former is that it exploits both local and global information.
The latter is that, by using some optimization, it efficiently provides good results.

This way, our approach is able to discover the community structure in, possibly large, networks. 
Our experimental evaluation, carried out over both synthetic and real-world networks, proves the efficiency and the robustness of the proposed strategy.
Some future directions of research include the creation of a friendship recommender systems which suggests new possible connections to the users of a Social Network, based on the communities they belong to.
Finally, we plan to design an algorithm to estimate the strength of ties between two social network users: for instance, in the case of networks like Facebook, this is equivalent to estimate the friendship degree between a pair of users.






\bibliographystyle{IEEEtran}

\bibliography{IEEEabrv,isda2011-k-path-bib}

\end{document}